\documentclass[prd,onecolumn,floatfix,preprintnumbers,showpacs,nofootinbib]{revtex4}

\usepackage{graphicx,color,natbib,enumerate}
\usepackage{dcolumn}% Align table columns on decimal point
\usepackage{bm}% bold math
\usepackage{epsfig}
\usepackage{hyperref}
\usepackage{amsmath} % Using 'align' in stead of 'eqnarray'.

\def\aj{\rm{AJ}}                   
             
\def\apj{\rm{ApJ}}                 
\def\apjl{\rm{ApJ}}                
\def\apjs{\rm{ApJS}}

\def\mnras{\rm{MNRAS}}

\def\prd{\rm{Phys.~Rev.~D}}        
        
\def\prl{\rm{Phys.~Rev.~Lett.}}

\def\lsim{\mathrel{\mathop
  {\hbox{\lower0.5ex\hbox{$\sim$}\kern-0.8em\lower-0.7ex\hbox{$<$}}}}}
\def\gsim{\mathrel{\mathop
  {\hbox{\lower0.5ex\hbox{$\sim$}\kern-0.8em\lower-0.7ex\hbox{$>$}}}}}

\newcommand{\nc}{\newcommand}

\nc{\be}[1]{\begin{equation}\mbox{$\label{#1}$}}
\nc{\bea}[1]{\begin{eqnarray} \mbox{$\label{#1}$}}
\nc{\Section}[2]{\section{#2}\label{#1}}
\nc{\Bibitem}[1]{\bibitem{#1}}
\nc{\Label}[1]{\label{#1}}
\nc{\Mpc}{Mpc/h}
\nc{\vev}[1]{\langle #1 \rangle}
\nc{\eea}{\end{eqnarray}}
\nc{\ee}{\end{equation}}

\begin{document}

\title{Cosmological forecasts from photometric measurements of the  
angular correlation function}

\author{F. Sobreira$^{a,b}$, F. de Simoni$^{b,c,d}$, 
R. Rosenfeld$^{a,b}$, L. A. N. da Costa$^{b,c}$, 
M. A. G. Maia$^{b,c}$ and M. Makler$^{b,e}$}
%\email{sobreira@ift.unesp.br}
\affiliation{$^a$Instituto de F\'{i}sica Te\'orica, Universidade Estadual Paulista, 
Rua Dr. Bento T. Ferraz, 271, S\~ao Paulo, SP 01140-070, Brazil \\
$^b$Laborat\'{o}rio Interinstitucional de e-Astronomia- LineA, Rua Gal. Jos\'{e} Cristino 77, 
Rio de Janeiro, RJ 20921-400,, Brazil \\
$^c$Observat\'orio Nacional, Rua Gal. Jos\'{e} Cristino 77, Rio de Janeiro, RJ 20921-400, Brazil\\
$^d$ 
Departamento de F\'{i}sica e Matem\'{a}tica, PURO/Universidade Federal Fluminense, Rua Recife s/n, Rio das Ostras, RJ 28890-000, Brazil\\
$^e$
Centro Brasileiro de Pesquisas F\'{\i}sicas, Rua Dr.\ Xavier Sigaud, 150, Rio de Janeiro, RJ 22290-180, Brazil
}

\date{\today}

\pacs{95.36.+x, 98.65.Dx, 98.80.Es}
\begin{abstract}
We study forecasts for the accuracy of the determination of 
cosmological parameters from future large scale photometric surveys 
obtained using the full shape of the 2-point galaxy angular correlation function.
The effects of linear redshift-space distortion, photometric redshift gaussian errors, 
galaxy bias and non-linearities in the power spectrum are included on our analysis.
The Fisher information matrix is constructed with the full covariance matrix, 
including the correlation between nearby redshift shells arising
from the photometric redshift error. 
We show that under some reasonable assumptions, a survey such as the imminent 
Dark Energy Survey should be able to 
constrain the dark energy equation of state parameter $w$ and
the cold dark matter density $\Omega_{cdm}$  with a precison of 
the order of $20 \%$ and $13 \%$ respectively  
from the full shape of the angular correlation function alone. 
When combined with priors from other observations the precision in the
determination of these parameters improve to $8 \%$ and $4 \%$ respectively.
\end{abstract}

\maketitle

\section{Introduction}
%\label{intro}

The clustering of large-scale structure of the universe is an invaluable source of 
information about the fundamental parameters that determine a given cosmological model. 
In the standard cosmological model, its origin is related to quantum fluctuations 
of the inflaton field during the 
inflationary period, and its evolution is determined by the different components 
of the universe that contribute to the total energy-momentum tensor, 
such as cold dark matter, dark energy, baryons and relativistic particles.

Until recently, cosmological constraints from galaxy distribution were obtained 
with spectroscopic catalogues 
mostly from the 2dF \cite{2dFref} and SDSS \cite{sdssref} projects. 
These projects raised the study of the galaxy clustering in 3D space to a new level, 
with a main focus on 
the 2-point correlation function, both in configuration and Fourier space. 
In particular, they resulted in the first observation of a feature in the 
correlation of the matter ditribution, 
arising from the so-called baryon acoustic oscillations (BAO) that occurred at an 
early stage of the evolution of the universe \cite{Eisenstein}. 

In the near future, projects such as the 
WiggleZ Dark Energy Survey \footnote{See http://wigglez.swin.edu.au} and the
Baryon Oscillation Spectroscopic Survey\footnote{See http://cosmology.lbl.gov/BOSS} (BOSS)  
will continue improving the spectroscopic catalogues. 
Future projects, both ground based such as BigBOSS\footnote{See http://bigboss.lbl.gov} and
space based such as Euclid\footnote{See http://sci.esa.int/euclid} will also rely on spectroscopic 
measurements of objects.

In order to collect a larger number of galaxies more efficiently, 
there are a number of planned galaxy surveys that will estimate the galaxy redshift 
from broad-band photometry measurements (photometric redshift or photo-z) with the 
aim to measure more than hundreds of millions up to a few billions of galaxies 
in a larger volume compared to spectroscopic surveys. 
The idea is to have a 
trade-off between precise spectroscopic redshifts for a
relatively small number of galaxies and less precise photometric redshifts for a
larger number of objects.
For instance,  the MegaZ-LRG  photometric catalogue based on the SDSS-II DR7 has become
recently avaliable \cite{CollisterEtAl,MegaZ}.
The measurement of the angular correlation function (ACF) for a 1.5 million 
Luminous Red Galaxies (LRG) with photometric redshifts 
was perfomed in \cite{Sawangwit09}, whereas their
angular power spectra (APS) was measured in \cite{MegaZ}.

The Dark Energy Survey\footnote{See http://www.darkenergysurvey.org} (DES) is a large dedicated
photometric survey that will soon start observations.
It is expected to measure ${\cal O} (300$ million) galaxies over an area of $5000$ 
square degrees of the southern sky
using 525 nights in the Blanco 4-meter telescope in Chile.
The main goal of DES is the measurement of the dark energy equation of state parameter $w$ up to redshift $z\sim 1.4$.  
This measurement will be performed using four complementary methods: 
distances from supernovae, large scale structure from the galaxy distribution, 
number count of galaxy cluster and shear from weak leasing measurements \cite{DESwhite}.     

The purpose of this paper is to determine forecasts for the precision 
of the determination cosmological parameters from
future photometric measurements of the large scale structure of the universe, such as what 
will be obtained by DES, using for the first time 
the full shape of the galaxy correlation function.
The use of the angular correlation function in a given redshift shell as an 
observable to infer cosmological parameters can be more appropriate than the use
of the spatial correlation function in the presence of errors inherent in the 
photometric measurements of redshifts. 
We use the full shape of the angular correlation function and its covariance, both at
different angles and in different redshift shells, to perform a Fisher matrix analysis of 
the sensitivity to cosmological parameters.  
Hence, we do not need to assume any parametric form of the ACF, as done in
an interesting recent study \cite{simoni}. Also our analysis is complementary to 
the study of the angular power spectrum since both techniques have 
advantages and disadvantages. The main disadvantage of the ACF is that the covariance
matrix is highly degenerate due to large correlations among different angular scales
whereas for the power spectrum errors in a given angular scale may propagate 
into several peaks.

In this paper we model the ACF with
6 cosmological parameters, assuming a spatially flat universe:  
the dark matter density parameter
$\Omega_{cdm}$, the baryon density parameter $\Omega_{b}$,
the Hubble parameter $h$, the primordial index of scalar perturbations  $n_{s}$,
the normalization of perturbations $\sigma_8$ and the equation of
state parameter for dark energy $w$.
In addition we allow for different
bias between dark matter and galaxies in each of the redshift shells. 
We include in the modelling the effects of 
linear redshift-space distortion, photometric redshift gaussian errors, galaxy bias and 
non-linearities in the power spectrum. 
The Fisher information matrix is constructed with the full covariance matrix from the model. 
In particular, we take into account the correlation between nearby redshift shells arising
from the photometric redshift error. We then discuss various forecasts for the cosmological
parameters in different scenarios.

This paper is organized as follows. In the next Section we present
the model for the angular correlation function. We include the effects
of nonlinearities in the power spectrum, redshift distortions from peculiar velocities 
of galaxies and the photo-z errors. 
In order to develop some intuition, we discuss 
in the third Section how the cosmological parameters affect the ACF.
Section IV is devoted to the modelling of the covariance matrix for the measurement
of the ACF in different redshift shells with different angular binnings. The Fisher matrix 
approach is discussed in Section V. Our main results are presented in Section VI and 
Section VII provides a summary and our conclusions.

\section{Modelling the angular correlation function}

The angular correlation function $\omega(\theta)$ is related to the two-point spatial 
correlation function $\xi^{(s)}$ in redshift space by
\begin{equation}
\omega(\theta)=\int_0^\infty dz_1 f(z_{1}) \int_0^\infty dz_2 f(z_{2}) \xi^{(s)} \left( r(z_1,z_2,\theta)\right).
\label{angular}
\end{equation}
The function $f(z)$ is determined by the selection function of the survey 
$\phi(z)$, the dark matter to luminous bias factor $b(z)$ 
and the linear growth function $D(z)$ (normalized to $D(z=0) =1$) 
as  $f(z) = \phi(z) b(z) D(z)$.
The radial comoving distance $r(z_1,z_2,\theta)$ is the distance between two galaxies at 
redshifts $z_1$ and $z_2$ separated by an angle $\theta$ 
and in a flat cosmology (which we will assume here) it is given by the relation, 
\begin{equation}
r = \sqrt{\chi^2(z_1)+\chi^2(z_2)-2\chi(z_1)\chi(z_2)\cos\theta},
\label{dist}
\end{equation}
where $\chi(z_i)$ is the comoving distance of the object $i$ to us (hereafter we use units with $c=1$):
\begin{equation}
\chi(z)=  \int_0^z\frac{dz'}{H(z')}
\end{equation}
and $H(z)$ is the usual Hubble function determined by the composition of the universe.

The redshift-space spatial correlation function $\xi^{(s)}$ 
corrected from the effects of redshift distortions
arising from the peculiar velocities of galaxies is given by (in the plane-parallel approximation)
\cite{Hamilton,Matsubara2000}
\begin{eqnarray}
\xi^{(s)}(r) &=&  \left[ 1 + \frac{1}{3} \left[ \beta(z_{1}) + \beta(z_{2}) \right] + \frac{1}{5} \beta(z_{1})\beta(z_{2}) \right] \xi_{0}(r)P_{0}(\mu) \nonumber \\ 
 &-& \left[ \frac{2}{3} \left[ \beta(z_{1}) + \beta(z_{2}) \right]
 + \frac{4}{7} \beta(z_1) \beta(z_2) \right] \xi_{2}(r)P_{2}(\mu) \nonumber \\ 
 &+& \left[  \frac{8}{35}\beta(z_{1})\beta(z_{2})\right] \xi_{4}(r)P_{4}(\mu) . 
\end{eqnarray}
Here the $P_{\ell}(\mu)$ are the usual Legendre polynomials as a function 
of $\mu=\hat{d} \cdot \hat{r}$ (cosine of angle between the line of sight $d$ and
$r$) and
$\beta(z) = g(z)/b(z)$ with $g(z) = d \ln D/d \ln a$.
The correlation multipoles are related to the matter power-spectrum $P_m(k)$ through:
\begin{equation}
\xi_{l}(r) = \frac{1}{2\pi^{2}} \int dk k^{2} P_m(k) j_{l}(kr)
\end{equation}

The selection function $\phi(z)$ is normalized such that:
\begin{equation}
\int_0^\infty dz \; \phi(z) = N
\end{equation}
where $N$ is the total number of objects per unit solid angle of the survey.
One usually slices the survey into $n$ redshift bins $i$ in such a way that:
\begin{equation}
\phi(z) = \sum_i \phi_i(z),
\end{equation}
with
\begin{equation}
\phi_i(z) = n(z) W_i(z),
\label{phi}
\end{equation}  
where $n(z)$ is the number density of galaxies per unit solid angle and per unit redshift and $W_i(z)$ is
a window function that selects the $i$-th redshift bin.
In general, $n(z)$ is a function of the limiting
magnitude of the survey.  We will adopt
a function of the form \cite{selection}
\begin{equation}
n(z) \propto (z/\bar{z})^2 e^{-\left( z/\bar{z} \right)^{1.5}}
\end{equation}
with $\bar{z}=0.5$ being the median redshift of the survey and
\begin{equation}
W_i(z) = \Theta\left(z-z_i^{\mbox{\tiny low}}\right) \Theta\left(z_i^{\mbox{\tiny high}} - z\right),
\end{equation}
where $\Theta(x)$ is the Heaviside (or step) function.

However, in surveys where only photometric redshifts are available one should 
incorporate into the
selection function the probability 
$P\left(z^{\mbox{\tiny ph}}|z \right)$ of obtaining a true redshift $z$ given that a photometric redshift
$z^{\mbox{\tiny ph}}$ is measured \cite{Ma06,sun09,hearin10,CrocceEtAl}:
\begin{eqnarray}
\phi_i(z) &=&  n(z) \int dz^{\mbox{\tiny ph}} \; W_i\left( z^{\mbox{\tiny ph}} \right)
P\left(z^{\mbox{\tiny ph}}|z \right) = \\ \nonumber
&n(z)& \int_{z_i^{\mbox{\tiny low}}}^{z_i^{\mbox{\tiny high}}} dz^{\mbox{\tiny ph}} \; 
P\left(z^{\mbox{\tiny ph}}|z \right). 
\end{eqnarray}  
Of course in the case $P(z_{\mbox{\tiny ph}} | z) = \delta(z_{\mbox{\tiny ph}}-z)$, one
recovers Eq.(\ref{phi}).

Usually the probability function for a spectroscopically calibrated galaxies is written as a gaussian
distribution:
\begin{equation}
P_G\left(z^{\mbox{\tiny ph}}|z \right) = \frac{1}{\sqrt{2 \pi} \sigma_z} 
\exp\left[-\frac{\left(z - z^{\mbox{\tiny ph}} - z_{\mbox{\tiny bias}} \right)^2}{2 \sigma_z^2} \right],
\label{gaussian}
\end{equation}
with the error given by
\begin{equation}
\sigma_z(z) = \delta_\sigma (1+z).
\end{equation}
When not mentioned, we set $\delta_{\sigma}=0.03$ which is expected for 
LRGs samples \cite{CollisterEtAl,RossEtAl}.
We will also consider here an unbiased photo-z and set $z_{\mbox{\tiny bias}} = 0$.

Another effect that must be taken into account is the decrease in the power of the ACF
due to the nonlinear gravitational clustering \cite{crocce08}. 
We will follow Crocce et al \cite{CrocceEtAl} and model this effect phenomenologically by introducing
a nonlinear power spectrum $P_{NL}(k)$ and substituting 
(neglecting the so-called mode-coupling term):
\begin{equation}
P_m(k) \rightarrow P_{NL}(k) = P_m(k) \exp\left[-r_{NL}^2 k^2 D^2(z)/2 \right]
\end{equation}
with $r_{NL} \approx 7 $ Mpc h$^{-1}$. Even though this is a simple-minded approach, 
it is robust depending on the scale someone is probing. 
Crocce et al \cite{CrocceEtAl} showed that this approach is in good agreement with simulations 
above $40 h^{-1}$Mpc, 
therefore our analysis will be valid and applied above this scale.

\section{Cosmological information in the angular correlation function}
Having a model for the angular correlation function
one can study the dependencies on the cosmological parameters.
The model is specified by the usual cosmological parameters.
Throughout this paper we assume as fiducial cosmological model
a flat $\Lambda$CDM universe with parameters as determined by 
WMAP7 \cite{wmap7}: dark matter density parameter 
$\Omega_{cdm}= 0.222$, baryon density parameter $\Omega_{b} = 0.0449$,
Hubble parameter $h=0.71$, primordial index of scalar perturbations  $n_{s}=0.963$,
and normalization of perturbations $\sigma_8=0.801$.
In addition, we set $r_{NL} = 6.6 $ Mpc h$^{-1}$, $\delta_{\sigma}=0.03$ and $b=2$ 
\cite{BlakeEtAl}.

The cosmological parameters can affect the angular correlation function in four different ways: 
in the primordial matter power-spectrum $P_m(k)$, the growth function $D(z)$, 
the linear redshift-space distortion $\beta(z)$ and in the definition of comoving distances (geometry).

The primordial power spectrum is characterized by the parameters $\Omega_{m}$, $\Omega_{b}$, $h$, $n_{s}$ and $\sigma_8$ (the dark energy equation of state parameter $w$ has a negligible 
impact on the primordial power spectrum in most cases). 
The growth function depends on $\Omega_{m}$ and $w$. 
The redshift space distortion parameter depends mainly on $\Omega_{m}$,  and the comoving 
distances are determined by $\Omega_m$ and $w$.

It is well known that there are several degeneracies among the 
parameters and functions, such as bias, $\sigma_8$ and
the growth function. 
We will show below that the analysis of multiple redshift shells including 
redshift-space distortion can ameliorate this problem \cite{okumura}.

\begin{figure}[!tbh]
  \begin{center}
    \includegraphics[width=8.5truecm]{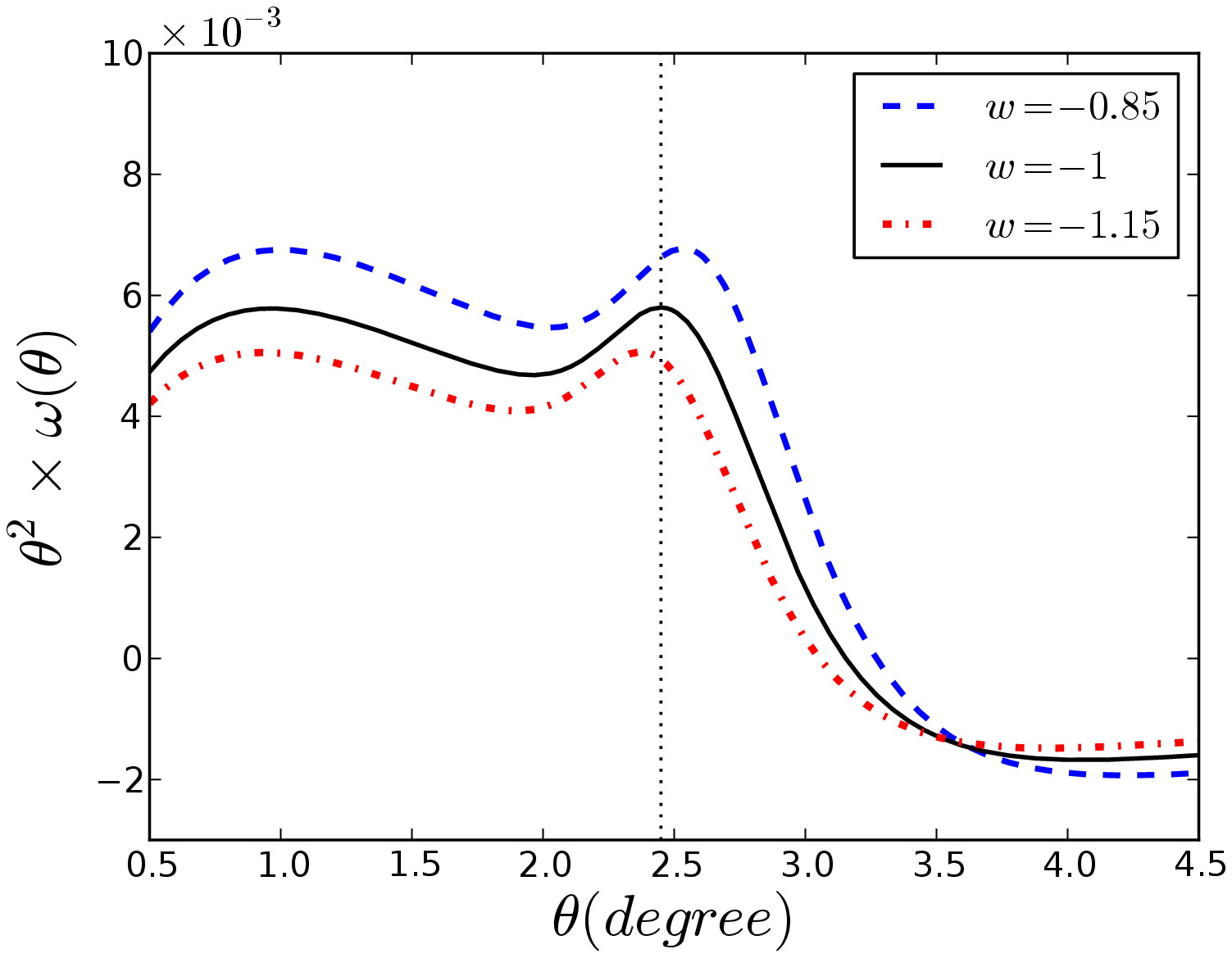}
\includegraphics[width=8.5truecm]{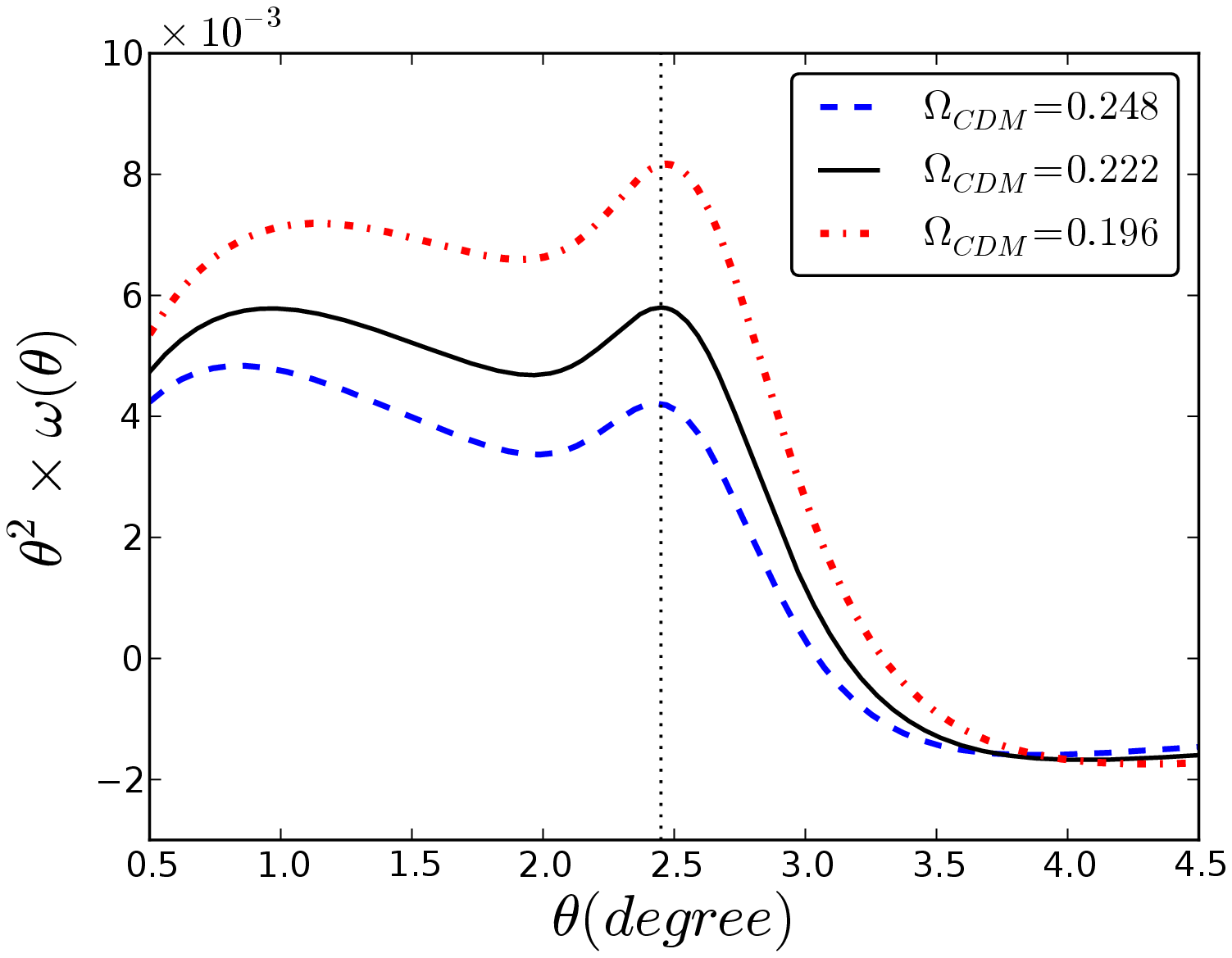}
  \end{center}
  \caption{\small Angular correlation function as a function of the angular scale 
for different values of the dark energy equation of state parameter 
$w$ (top panel) and the cold dark matter density parameter $\Omega_{CDM}$ (bottom panel) 
for the redshift 
shell $1.00 \leq z \leq 1.05$ with photo-z dispersion $\delta_{\sigma}=0.03$. 
The vertical dotted line indicates the peak position of $\theta^{2}\times \, \omega(\theta)$ 
in the fiducial cosmology as stated in the text.}
  \label{acf_w_cdm}
\end{figure}

In Fig.(\ref{acf_w_cdm}) we show how the angular correlation function changes with $w$ and $\Omega_{CDM}$ 
around our fiducial cosmology and for the redshift shell $1.0 \leq z \leq 1.05$. 
From the top panel one can see that the ACF changes mostly its amplitude and the location of the BAO feature. 
As stated above,
the primordial power spectrum does not change appreciably with $w$, but the growth factor 
$D(z)$ and the comoving distance are affected by this parameter. 
The contribution from the growth factor can be easily estimated.
Since for this shell $D(w=-1) = 0.618$ and $D(w=-0.85) = 0.630$, the growth function 
can account for an
increase of $\sim 3.9\%$ in the amplitude. Hence it fall short of explaining the 
 $\sim 14\%$ increase of the amplitude at intermediate scales. 
Furthermore, the increase in $D(z)$
can not explain the shift in the position of the BAO feature.

It is noticeable that the BAO feature shifts towards larger angular scale when going from the fiducial cosmology $w=-1$ to $w=-0.85$, while when $w=-1.15$ it shifts to smaller angular scales. 
The explanation comes from the fact that one changes the geometry when changing $w$, 
and distances have different values, changing the ''standard ruler". 
When converting the spatial correlation function to find the ACF, 
the projection depends on the thickeness of the shell in comoving distance. 
If the shell is wider (in comoving distance) the ACF will have less power and the 
projection offset will be higher. 
For the fiducial cosmology the thickeness of the shell $1.0\leq z \leq 1.05$ 
when converted  to comoving distance is $\Delta \chi \simeq 87.4 \, h^{-1}$Mpc, while for $w=-0.85$ we have  $\Delta \chi \simeq 83.6 \, h^{-1}$Mpc, resulting in a modification 
of $\sim 5\%$ in the thickness of the comoving shell. 
This fact also explains why the BAO feature shifts to smaller angles from the fiducial 
value to $w=-0.85$ \cite{simpson,simoni}. 
It is interesting that such a small change in the shell thickness 
can generate a non-negligible modification in the ACF shape and amplitude.

The parameter $\Omega_{CDM}$ has an impact in all four effects listed above and its impact on
the ACF is shown in the bottom panel of Fig.(\ref{acf_w_cdm}).
However, most of the cosmological information for $\Omega_{CDM}$ will come from the primordial power spectrum,
with modifications  in the 
fraction of baryons $f_{b} = \Omega_{b}/\Omega_{m}$ and the product $\Omega_{m}h$. 

A complication in the simple effects illustrated above is the presence of photo-z errors, 
which can mimic some of the effects of changing the cosmological parameters, in addition
to redshift space distortions, nonlinearities and galaxy bias.
All these effects are of course taken into account in our results.

\section{Modelling the errors in the calculation of the ACF}

In order to derive meaningful cosmological constraints or even determine
the reach of a given survey, one has to 
understand the errors involved in the measurements of the ACF. More precisely,
one must model the full covariance matrix for measurements of the ACF 
at different angles and in different shells.

\subsection{Covariance matrix for one shell}

We will follow closely the works of Blake el al \cite{BlakeEtAl} and Crocce et al \cite{CrocceEtAl}
in our modelling. The modelling will include effects from photo-z errors, redshift distortion, 
partial sky coverage and shot noise. For this analysis it is more convenient to work with the angular power spectrum (a derivation of the ACF covariance matrix in configuration space 
can be found in Cohn \cite{cohn06}). 
We define the projected density fluctuations onto the sky in a particular direction $\hat{n}$, 
$\sigma(\hat{n})$ as
\begin{equation}
\sigma(\hat{n}) = \int dz \; \phi(z) \; \delta(\hat{n},z)
\label{sigma}
\end{equation}
and decompose it in spherical harmonics:
\begin{equation}
\sigma(\hat{n}) = \sum_{l=0}^{\infty} \sum_{m=-l}^{l} a_{lm} Y_{lm}(\hat{n}).
\label{spherical}
\end{equation}
The angular correlation function can be written as:
\begin{eqnarray}
\omega(\theta) &=& \langle \sigma\left(\hat{n}\right) \sigma\left(\hat{n} + \vec{\theta} \right) \rangle 
\\ \nonumber
&=& 
\langle \sum_{lm} \sum_{l' m'} a_{lm} a_{l'm'} Y_{lm}(\hat{n}) Y_{l'm'}(\hat{n} + \vec{\theta}) \rangle .
\end{eqnarray}
From isotropy we can define $C_l$ from
\begin{equation}
\langle a_{lm} a_{l'm'} \rangle = \delta_{l l'} \delta_{m m'} C_l
\end{equation}
and therefore
\begin{equation}
\omega(\theta) = \sum_{lm}  C_l  Y_{lm}(\hat{n}) Y_{lm}(\hat{n} + \vec{\theta}) = 
\sum_l C_l \frac{2l+1}{4 \pi} P_l(\cos \theta),
\label{ACFfromCl}
\end{equation}
where the addition theorem was used in the last equality.

The $C_l$'s are the variance of the distribution for the coefficients $a_{lm}$, assumed gaussian:
\begin{equation}
P(a_{lm})da_{lm} = \frac{1}{\sqrt{2 \pi C_l}} e^{-\frac{a_{lm}^2}{2 C_l}} da_{lm}
\end{equation}

The covariance matrix is given by:
\begin{eqnarray}
&&\mbox{Cov} (\omega(\theta) \omega(\theta') ) = \langle \omega(\theta) \omega(\theta') \rangle = 
\\ 
&&\sum_{ll'} \mbox{Cov}(C_l C_{l'}) \frac{2l+1}{4 \pi} \frac{2l'+1}{4 \pi} P_{l}(\cos \theta) P_{l'}(\cos \theta')
\nonumber
\end{eqnarray}

For full sky the $C_l$'s are statistically independent. For partial sky coverage one can use an averaged $C_l$
over a band of width $\Delta l =10$ and these will be independent to a good approximation. We will assume that
\begin{equation}
\mbox{Cov}(C_l C_{l'}) = \mbox{Var}(C_l) \delta_{l l'}
\end{equation}
and assuming a gaussian distribution for the likelihood function of $C_l$ \cite{Dodelson},
correcting for partial sky coverage $f_{\mbox{\tiny sky}}$ and including shot-noise error one has:
\begin{equation}
\mbox{Var}(C_l) = \frac{2}{(2 l+1) f_{\mbox{\tiny sky}}} \left( C_l + 1/\bar{n} \right)^2
\end{equation}
where $\bar{n} = N/\Delta \Omega$ is the average number of galaxies per unit solid angle.

Therefore the covariance matrix can be computed as:
\begin{equation}
\mbox{Cov} (\omega(\theta) \omega(\theta') ) = \frac{2}{f_{\mbox{\tiny sky}}}
\sum_{l}  \frac{2l+1}{(4 \pi)^2} P_{l}(\cos \theta) P_{l}(\cos \theta') \left( C_l + 1/\bar{n} \right)^2
\end{equation}

We can estimate the $C_l$'s from a model for the power spectrum by first performing a Fourier transform in 
the 3-d density field in eq.(\ref{sigma})
\begin{equation}
\delta(\hat{n},z) = \int \frac{d^3k}{(2 \pi)^3} \delta(\vec{k},z) e^{i \vec{k}\cdot \hat{n} r(z)}
\end{equation}
and use the identity
\begin{equation}
e^{i \vec{k} \cdot \hat{n} r} = 4 \pi \sum_{lm} i^l j_l(kr) Y_{lm}(\hat{n}) Y_{lm}^\ast(\hat{k})
\end{equation}
to compare with eq.(\ref{spherical}) and find:
\begin{equation}
a_{lm}^i = \int dz \; \phi_i(z) \; \int \frac{d^3k}{(2 \pi)^3} \delta(\vec{k},z) 4 \pi i^l  j_l(kr) Y_{lm}^\ast(\hat{k})
\end{equation}
 
Defining
\begin{equation}
\Psi_l^i(k) = \int dz \; \phi_i(z) D(z) j_l(k r(z)) b(z)
\end{equation}
with  $\delta(\vec{k},z) = D(z) \delta(\vec{k},0)$
allows us to write:
\begin{equation}
C_l^i = \langle |a_{lm}^i|^2 \rangle = \frac{2}{\pi} \int dk\; k^2 P_{NL}(k) \left(\Psi_l^i\right)^2(k)
\label{cl}
\end{equation}

At this point one must also introduce the effects of redshift distortions, 
as done for the angular correlation function.
We will adopt the prescription given in \cite{PaddyEtAl,CrocceEtAl} and add to the function $\Psi_l(k)$ another term $\Psi^r_l(k)$
that incorporates the redshift distortion given by:
\begin{eqnarray}
&\Psi^{i,r}_l(k)& =  \int dz \; \beta(z) \phi_i(z) D(z) \left[ \frac{2l^2+2l-1}{(2l+3)(2l-1)} j_l(kr) 
\right.  \\
&-& \left.\frac{l(l-1)}{(2l-1)(2l+1)} j_{l-2}(kr) - \frac{(l+1)(l+2)}{2l+1)(2l+3)} j_{l+2}(kr) \right].
\nonumber
\end{eqnarray}

\begin{figure}[!tbh]
  \begin{center}
    \includegraphics[width=8.5truecm]{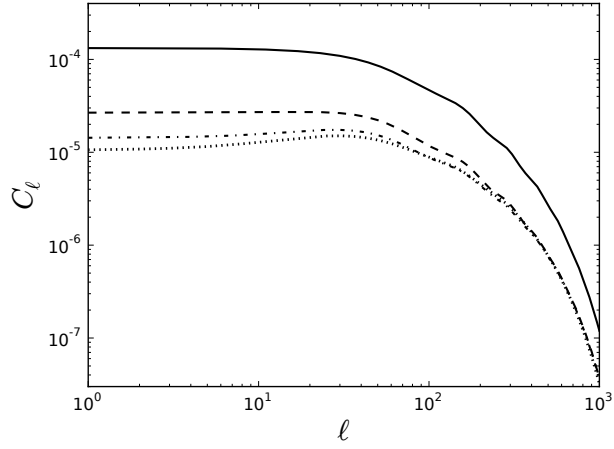}
  \end{center}
  \caption{\small Photo-z error impact on the angular power spectrum for the redshift shell $0.95\leq z\leq 1.00$. The solid line correspond to no errors in the redshift determination, dashed line with $\delta_{\sigma}=0.03$, dot-dashed line $\delta_{\sigma}=0.05$ and dotted line $\delta_{\sigma}=0.10$.}
  \label{cl_photozimpact}
\end{figure}

In Fig. (\ref{cl_photozimpact}) we show the impact of the photo-z error on the angular power spectrum for $\delta_{\sigma}=0.03, 0.05$ and $0.10$. One can see how the
power decreases with increasing photo-z error, as expected.  
We also have checked the numerical agreement of the angular correlation function computed from
eq.(\ref{ACFfromCl}) with the sum on spherical harmonics up to $l=1000$ 
(to ensure numerical accuracy) with 
the definition given in eq.(\ref{angular}).

\subsection{Covariance matrix including different shells}

The various redshift bins of the survey are not independent due to the
errors in the photometric redshift. Of course the main
correlations will occur for adjacent bins. 
Hence we need covariance matrices for 2 different redshifts $i$ and $j$ \cite{BlakeEtAl}:
\begin{eqnarray}
&& \mbox{Cov}(\omega^i(\theta_n) \omega^j(\theta_m) )  = \langle \omega^i(\theta_n) \omega^j(\theta_m) \rangle 
\\
&=& 
\sum_{ll'} \mbox{cov}(C_l^i C_{l'}^{j}) \frac{2l+1}{4 \pi} \frac{2l'+1}{4 \pi} P_{l}(\cos \theta_n) P_{l'}(\cos \theta_m) \nonumber
\end{eqnarray}
where in this case, since the shot noise between shells are uncorrelated, we write
\begin{equation}
\mbox{Cov}(C_l^i C_{l'}^j) = \frac{2}{(2 l+1) f_{\mbox{\tiny sky}}} \left(C^{i,j}_l + 1/\bar{n}_i \; \delta_{ij}\right)^2 \delta_{l l'}
\end{equation}

Therefore, in the general multibin case the full covariance matrix can be computed as:
\begin{eqnarray}
 \mbox{Cov}(\omega^i(\theta_n) \omega^j(\theta_m) ) = 
\frac{2}{f_{\mbox{\tiny sky}}}
\sum_{l} && \left[ \frac{2l+1}{(4 \pi)^2} P_{l}(\cos \theta_n) \right.\\ \nonumber  
&& \left. P_{l}(\cos \theta_m) \left( C_l^{i,j} + 1/\bar{n}_i \; \delta_{ij} \right)^2 \right] 
\end{eqnarray}
with
\begin{equation}
C_l^{i,j} = \frac{2}{\pi} \int dk\; k^2 P(k) \Psi_l^i(k) \Psi_l^j(k).
\label{cl}
\end{equation}

\begin{figure}[!tbh]
  \begin{center}
    \includegraphics[width=8.5truecm]{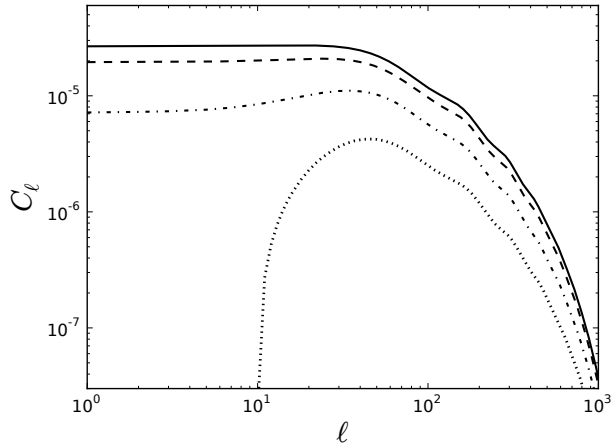}
  \end{center}
  \caption{\small Cross-correlation angular power spectrum for the redshift shell $0.95\leq z\leq 1.00$ with its 3 next neighbours. 
The solid line correspond to auto-correlation. The photo-z error was set to $\delta_{\sigma}=0.03$.}
  \label{cl_cross}
\end{figure}

In Fig(\ref{cl_cross}) it is shown the cross-correlation angular power spectrum $C_l^{i,j}$
for the redshift shell $0.95\leq z\leq 1.00$ with itself (auto-correlation) and its 3 nearest neighbours on one side. 
One can see that the cross-correlation for the nearest 2 shells is enough
to capture the relevant effect. Hence in the following we will include the cross-correlation
only with the neighbouring 4 shells, 2 on each side. 

\section{Error forecasts for cosmological parameters}

We will use the full shape of the angular correlation function $\omega(\theta)$ in 
our analysis to derive
forecasts for the errors in the determination of cosmological parameters 
using a Fisher matrix approach. 

The Fisher matrix approach approximates the full likelihood function of the parameters by a 
multivariate Gaussian distribution
of the parameters \cite{TTH,Dodelson}:
\begin{equation}
{\cal L }(\{p\}) \propto \exp \left[ -\frac{1}{2} (p_\alpha - \bar{p}_\alpha) F_{\alpha \beta} (p_\beta - \bar{p}_\beta) \right],
\end{equation}
that is, the Fisher matrix $F_{\alpha \beta}$ is the inverse of the covariance matrix for the parameters:
\begin{equation}
F_{\alpha \beta} = - \left[\frac{\partial^2 \log {\cal L}(\{p\}) }{\partial p_\alpha \partial p_\beta} \right]_{\{\bar{p}\}}
\end{equation}
where $\{\bar{p}\}$ denotes the best-fit parameters. We do not know a priori what 
the best-fit parameters are, since this involves a complicated search in the 
space with dimensions given by the total number of parameters. 
We will compare in the Appendix a few results from Fisher matrix 
with brute force computation of the likelihood function.

The most useful application of the Fisher matrix approach is to 
study {\it forecasts} of future experiments, determining the precision 
with which the parameters can be measured with respect to {\it fiducial} values.
These fiducial values replace the unknown best fit parameters.

The Fisher matrix in this case is given by:
\begin{eqnarray}
F_{\alpha \beta} &=& \frac{ \partial \omega_{\mbox{\tiny th}}^i(\theta^n,p)}{\partial p_\alpha} 
 \left[ C^{-1}\right]^{ij}_{nm} \frac{ \partial \omega_{\mbox{\tiny th}}^j(\theta^m,p)}{\partial p_\beta}  \\ \nonumber
&+& \frac{1}{2} \mbox{Tr} \left[  C^{-1} 
\frac{ \partial C}{\partial p_\alpha}  C^{-1} 
\frac{ \partial C}{\partial p_\beta} \right]
\label{fisher} 
\end{eqnarray}
where $C^{ij}_{nm} = \mbox{Cov}(\omega^i(\theta_n) \omega^j(\theta_m) )$ 
is the covariance matrix derived in the previous section and 
the derivatives are calculated at the fiducial set of parameters $\{\bar{p}\}$.
In general the first term dominates over the second one \cite{Tegmark97} and our results will
be obtained using only the first term. This is a conservative result and further below we
will show that the second term indeed has a small impact on the forecast.

\section{Results}

We have divided a DES-like survey into 20 redshift shells with width $\Delta z = 0.05$ 
in the range $0.4<z<1.4$. We justify this value of $\Delta z$ below. 
The angular binning varies with the redshift shell in order to select 
angles around the BAO peak position. 
The redshift and angular binnings are shown in Table (\ref{binning})
for the case of roughly $60<r<160$ $h^{-1}$Mpc.
The full correlation matrix in our case turns out to be a $445 \times 445$ matrix.

\begin{table}
\begin{tabular}{|c|c|c|c|c|}
\hline
redshift shell & angular range  & bins &bin width  & spatial scale  \\ 
\hline \hline
$0.40<z<0.45$ &3 - 8 &26 &0.2 &$60<r<160$ \\ \hline
$0.45<z<0.50$ &3 - 8 &26 &0.2 &$67<r<180$ \\ \hline
$0.50<z<0.55$ &2.5 - 6.1 &25 &0.15 &$61<r<149$ \\ \hline
$0.55<z<0.60$ &2.5 - 6.1 &25 &0.15 &$66<r<162$ \\ \hline
$0.60<z<0.65$ &2 - 5.6   &25 &0.15 &$57<r<159$ \\ \hline
$0.65<z<0.70$ &2 - 5.6   &25 &0.15 &$61<r<170$ \\ \hline
$0.70<z<0.75$ &1.7 - 5.0 &23 &0.15 &$55<r<162$ \\ \hline
$0.75<z<0.80$ &1.7 - 5.0 &23 &0.15 &$58<r<171$ \\ \hline
$0.80<z<0.85$ &1.5 - 4.5 &21 &0.15 &$54<r<161$ \\ \hline
$0.85<z<0.90$ &1.5 - 4.5 &21 &0.15 &$56<r<169$ \\ \hline
$0.90<z<0.95$ &1.5 - 4.5 &21 &0.15 &$59<r<177$ \\ \hline
$0.95<z<1.00$ &1.5 - 4.0 &21 &0.15 &$61<r<183$ \\ \hline
$1.00<z<1.05$ &1.3 - 3.7 &21 &0.12 &$55<r<157$ \\ \hline
$1.05<z<1.10$ &1.3 - 3.7 &21 &0.12 &$57<r<163$ \\ \hline
$1.10<z<1.15$ &1.3 - 3.7 &21 &0.12 &$59<r<168$ \\ \hline
$1.15<z<1.20$ &1.2 - 3.1 &20 &0.1 &$56<r<145$ \\ \hline
$1.20<z<1.25$ &1.2 - 3.1 &20 &0.1 &$58<r<150$ \\ \hline
$1.25<z<1.30$ &1.2 - 3.1 &20 &0.1 &$60<r<154$ \\ \hline
$1.30<z<1.35$ &1.2 - 3.1 &20 &0.1 &$61<r<158$ \\ \hline
$1.35<z<1.40$ &1.2 - 3.1 &20 &0.1 &$63<r<162$ \\ \hline
\end{tabular}
\caption{Binning in redshift and angles (in degrees) 
for roughly $60<r<160$ $h^{-1}$Mpc} 
\label{binning}
\end{table}

The $C_l$'s were computed in a suite of {\tt c} codes based on the 
free software Gnu Scientific Library \cite{gsl}. 
The evaluation of the integrals of the rapidly oscillating
spherical Bessel functions are particularly demanding.
We compute the  $C_l$'s up to $l=1000$ in order to obtain our results.
 
\begin{figure}[h!]
  \begin{center}
    \includegraphics[width=8.5truecm]{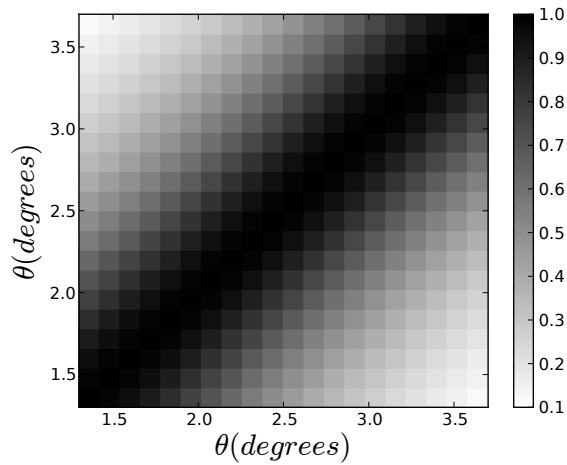}
  \end{center}
  \caption{\small Reduced covariance matrix for the bin $1.00<z<1.05$.}
   \label{Cov}
\end{figure}

The reduced covariance matrix for one bin ($1.00<z<1.05$), defined by 
$R_{nm} = C_{nm}/\sqrt{C_{nn} C_{mm}}$  is shown in Fig.(\ref{Cov}).
In this case we used $f_{\mbox{\tiny sky}} = 0.125$ and $\bar{n} = 15/\mbox{arcmin}^2$.
It can be seen that the errors in different angular bins are highly correlated resulting
in a quasi-singular covariance matrix. The magnitude of the first and second off-diagonal
elements is typically $0.9$ and $0.8$. This is one of the major disadvantages of
using the ACF to derive cosmological parameters.

In order to compute the Fisher matrix
we invert the covariance matrix using singular value decomposition \cite{numerical}.
%setting the elements of the diagonal matrix to zero if they are smaller than the 
%floating-point precision 

\subsection{Redshift shell width and shot noise}

\begin{figure}[h!]
  \begin{center}
    \includegraphics[width=3in]{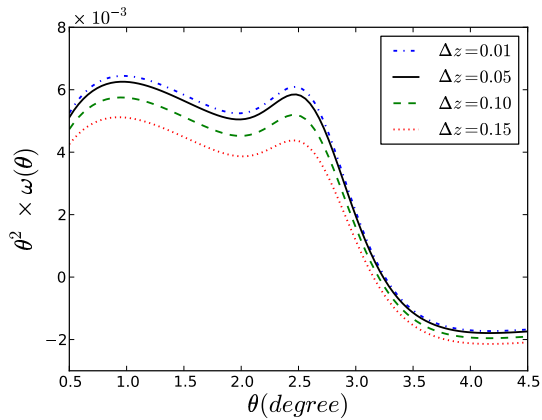}
  \end{center}
  \caption{\small Angular correlation function for a redshift bin centers at $z=1.1$ for 4 different
bin widths $\Delta z$. }
  \label{shell_width}
\end{figure}

The narrower the redshift shell is, the larger signal is obtained for the angular
correlation function, as illustrated in Fig.(\ref{shell_width}). However,
one must be concerned with the impact of shot-noise on the forecast of cosmological
parameters, which we analyze in this subsection. 

\begin{figure}[h!]
  \begin{center}
    \includegraphics[width=3in]{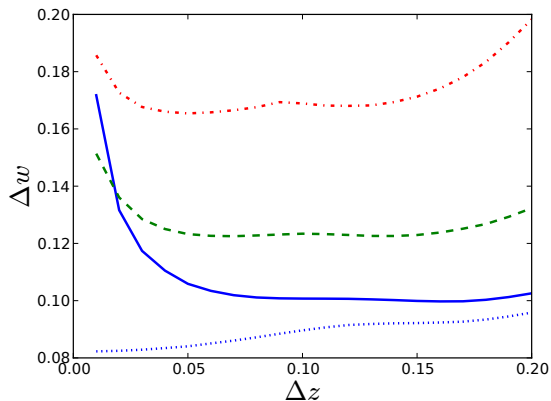}
  \end{center}
  \caption{\small 
The lines show the expected error on $w$ as a function of the width of the shells 
centered at 3 mean redshifts. For all of them we are assuming $N_{gal} = 300$ millions 
of galaxies and $\sigma_z = 0.03(1+z)$. 
The red dot-dashed line is centered at $z=0.9$, the dashed green for $z=1.1$ and 
the solid blue line for $z=1.3$. 
The dotted blue line is the same as the last shell but now neglecting 
the contribution from the shot-noise.}
  \label{shell_width_w}
\end{figure}

In Fig. (\ref{shell_width_w}) we plot the $1\sigma$ error forecast 
on the determination of the equation of state $w$ as a function of the shell width when 
marginalizing over the parameters $\Omega_{cdm}$, $\sigma_{8}$ and bias, with the other parameters fixed at their fiducial values, and
with a photo-z error given by $\sigma_{z}=0.03(1+z)$.
These forecasts were obtained for redshift shells centered at 3 
mean redshift with different shell widths $\Delta z$. 
As expected, one can see that for very narrow redshift shells
the shot-noise error becomes dominant even for the DES expected number of galaxies, 
which we assume to be $N_{gal} = 300$ million. 

The best value for the shell width is around $\Delta z \sim 0.05$ 
(a similar result was found by \cite{ross11}, 
where they found that the best shell width for the the redshift 
space distortion parameter is $\sim 0.07$). 
The blue dotted line shows the error forecast on $w$ when the shot-noise term is neglected. 
As expected if shot-noise is absent, 
the result improves as the width gets smaller because the 
angular correlation has more signal for thinner shells.

Another interesting feature of Fig.(\ref{shell_width_w}) is the flatness 
of the forecast error on 
$w$ for $\Delta z \gtrsim 0.05$. One naively would expect that the constraint should degrade for very wide shells due to 
the washing out of the spatial correlation function when projecting on the sphere. 
However, as pointed out by Simpson et al \cite{simpson}, the corresponding top-hat shell 
that simulates
a true redshift distribution, denoted by $\Delta z_{TH}$, can be obtained from the convolution of
the top-hat shell in photometric redshift space $\Delta z$ with the probability distribution 
for the photo-z error eq.(\ref{gaussian}), which can be approximated by
\begin{equation}\label{simpon_shell}
 \Delta z_{TH} = \sqrt{\Delta z^{2} + 12\sigma_{z}^{2}}\,\,.
\end{equation}
From this relation we find for $\Delta z = 0.01$, $0.05$, $0.10$ and $0.15$ the following top-hat values for the shell at 
$z = 1.1$ (green dashed line in figure \ref{shell_width_w}), 
respectively: 0.219, 0.224, 0.240 and 0.265. 
Therefore, only for very wide redshift shells the true (spectroscopic) top-hat shell will have an 
impact in the cosmological parameters due to projection effects compared to very thin shells. 
Hence in what follows we set $\Delta z = 0.05$ through the entire redshift range.

\subsection{Forecasts from the ACF}

\begin{figure}[h!]
  \begin{center}
    \includegraphics[width=3in]{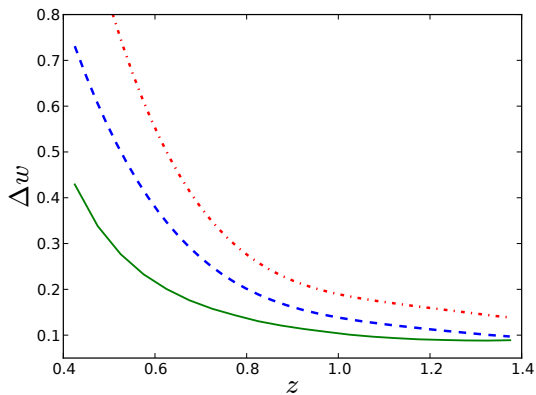}
  \end{center}
  \caption{\small Forecasts for 1-$\sigma$ errors on the equation of state $w$ as a function of redshift with marginalized bias, $\sigma_8$ and $\Omega_{cdm}$. 
Optimistic case with $\delta_\sigma = 0.03$ and $20<r<200$ $h^{-1}$Mpc (solid green line),
fiducial case with $\delta_\sigma = 0.03$ and $60<r<160$ $h^{-1}$Mpc (dashed blue line) 
and conservative case
$\delta_\sigma = 0.05$ and $60<r<160$ $h^{-1}$Mpc (dot-dashed red line). }
  \label{sensitivity}
\end{figure}

\begin{figure}[h!]
  \begin{center}
    \includegraphics[width=3in]{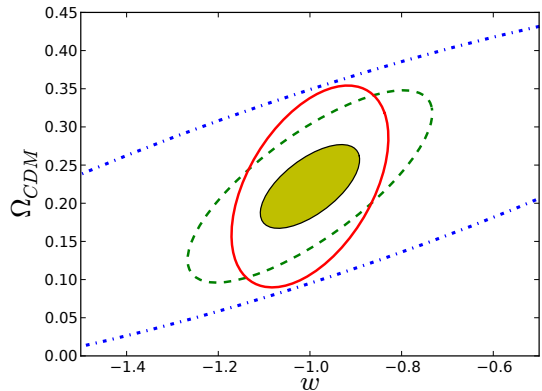}
  \end{center}
  \caption{\small Forecasts for $\Omega_{cdm}$  and $w$ from the angular correlation 
function of three redshift shells, 
$0.40<z<0.45$ (blue dot-dashed line),$0.85<z<0.90$ (green dashed line) and 
$1.35<z<1.40$ (red solid line). The yellow ellipse
is the result of combining the 3 shells. }
  \label{3shells}
\end{figure}

The sensitivity to the cosmological parameters for each shell increases with redshift, as exemplified in 
Fig.(\ref{sensitivity}), where we show the errors in the equation of state obtained from each shell independently.
This can also be seen in Fig.(\ref{3shells}), where we show 
the forecasts on the 
equation of state $w$ and the dark matter
density $\Omega_{cdm}$  obtained from 3 independent redshift shells, 
$0.40<z<0.45$, $0.85<z<0.90$ and $1.35<z<1.40$.
We have marginalized over $\sigma_8$ and bias (we allow for a free
bias parameter for each shell) 
but the other parameters are held at their fiducial values.
One notices that the forecasts are better for shells at larger 
redshifts and that the ellipses rotate slightly from one shell to the next, 
a behaviour which helps to break
some of the degeneracies. The forecast errors are $\sigma_w = 0.071$ and
$\sigma_{\Omega_{cdm}} = 0.036$.
 
\begin{figure}[h!]
  \begin{center}
    \includegraphics[width=3in]{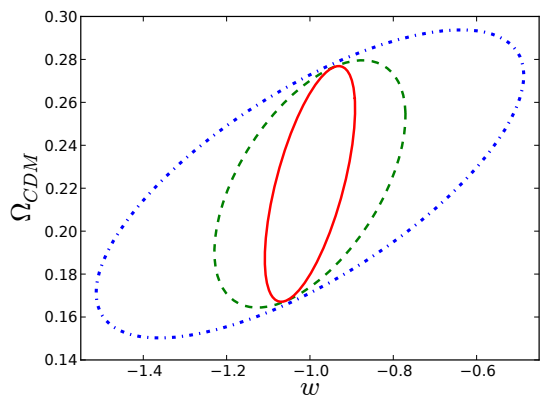}
  \end{center}
  \caption{\small Forecasts for $\Omega_{cdm}$  and $w$ from the angular 
correlation function of five redshift shells.
The red solid ellipse is the result with all parameters fixed at their 
fiducial values; blue dot-dashed ellipse is the
result with all parameters marginalized without priors; green dashed ellipse 
is the result with priors on $h$ and $\Omega_b$. }
  \label{allparameters}
\end{figure}

We also checked the robustness of this result if the other parameters are 
marginalized instead of fixed to their fiducial values. 
This is shown in Fig.(\ref{allparameters}), where the previous result with all 
parameters other than $\Omega_{cdm}$ and $w$ fixed
at their fiducial values is shown in the solid ellipse. 
As expected, marginalizing over all other parameters without any priors
(dot-dashed ellipse) degrades the forecasts significantly. 
The errors on the parameters without any priors are:
$\sigma_w=0.34, \sigma_{\Omega_{cdm}} = 0.047, \sigma_{\Omega_{b}} = 0.022, \sigma_h = 0.26,
\sigma_{n_s} = 0.31$ and $\sigma_{\sigma_8}= 0.77$.
However, introducing priors only on $h$ (from HST) and $\Omega_b$ from
WMAP7 (dashed ellipse) again reduces the errors of the forecasts.

\begin{figure}[h!]
  \begin{center}
    \includegraphics[width=3in]{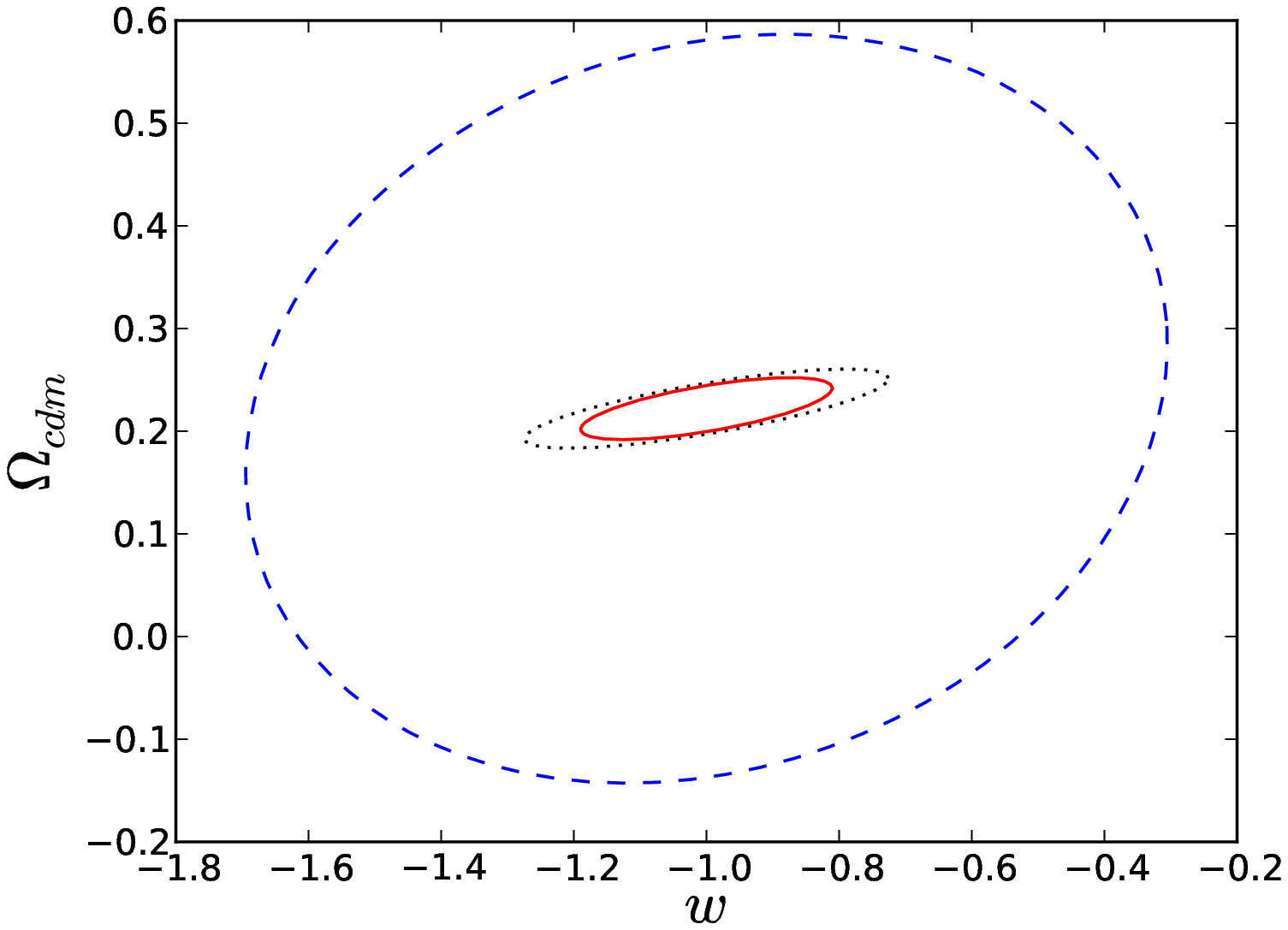}
  \end{center}
  \caption{\small Forecasts for $\Omega_{cdm}$  and $w$ from the angular correlation function for all redshift
shells without correlation among shells. Contribution from the neglected term 
(dashed line), contribution from
the dominant term (dotted line) and result from both terms (solid line).
 }
  \label{secondterm}
\end{figure}

As explained above, our results will be conservative since we are neglecting an extra 
term in the Fisher matrix. To assess the small impact 
of the neglected term, we plot in Fig.(\ref{secondterm}) the analysis 
for all redshift bins
without correlations among bins with and without the contribution 
from the neglected term. We see that, as expected, including the second
term leads to a slightly more restrictive forecast.

\begin{figure}[h!]
  \begin{center}
    \includegraphics[width=3in]{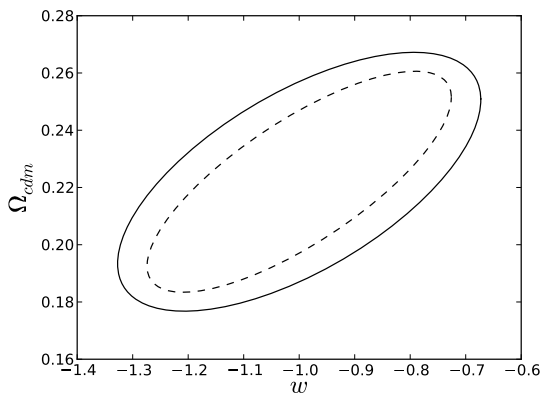}
  \end{center}
  \caption{\small Forecasts for $\Omega_{cdm}$  and $w$ from the angular correlation function with all redshift
shells included. All other parameters are marginalized with 
no priors and $\delta_\sigma=0.03$. Dashed line is the result with diagonal 
correlation matrix (only auto-correlations) and full line is the result taking 
into account next 2 neighbour bins.
}
  \label{diagXfull}
\end{figure}

We now proceed to our complete analysis. We consider 26 parameters, namely 
$ n_s, \sigma_8, h, \Omega_b, \Omega_{cdm}, w, b_1, b_2, \cdots, b_{20} $, 
where $b_i$ are the bias for the 
$i^{th}$ redshift shell. We include correlations for the adjacent 2 redshift bins, 
{\it i.e.} for each bin we take
correlations with 4 bins. In order to show the impact of including these correlations 
we show in
Fig.(\ref{diagXfull}) the 1$\sigma$ constraint on $\Omega_{cdm}$ and $w$ with all 
other parameters marginalized
without priors. As expected, it can be seen that the error 
is underestimated when the correlations 
are not taken into account.

\begin{figure}[h!]
  \begin{center}
    \includegraphics[width=5.0in]{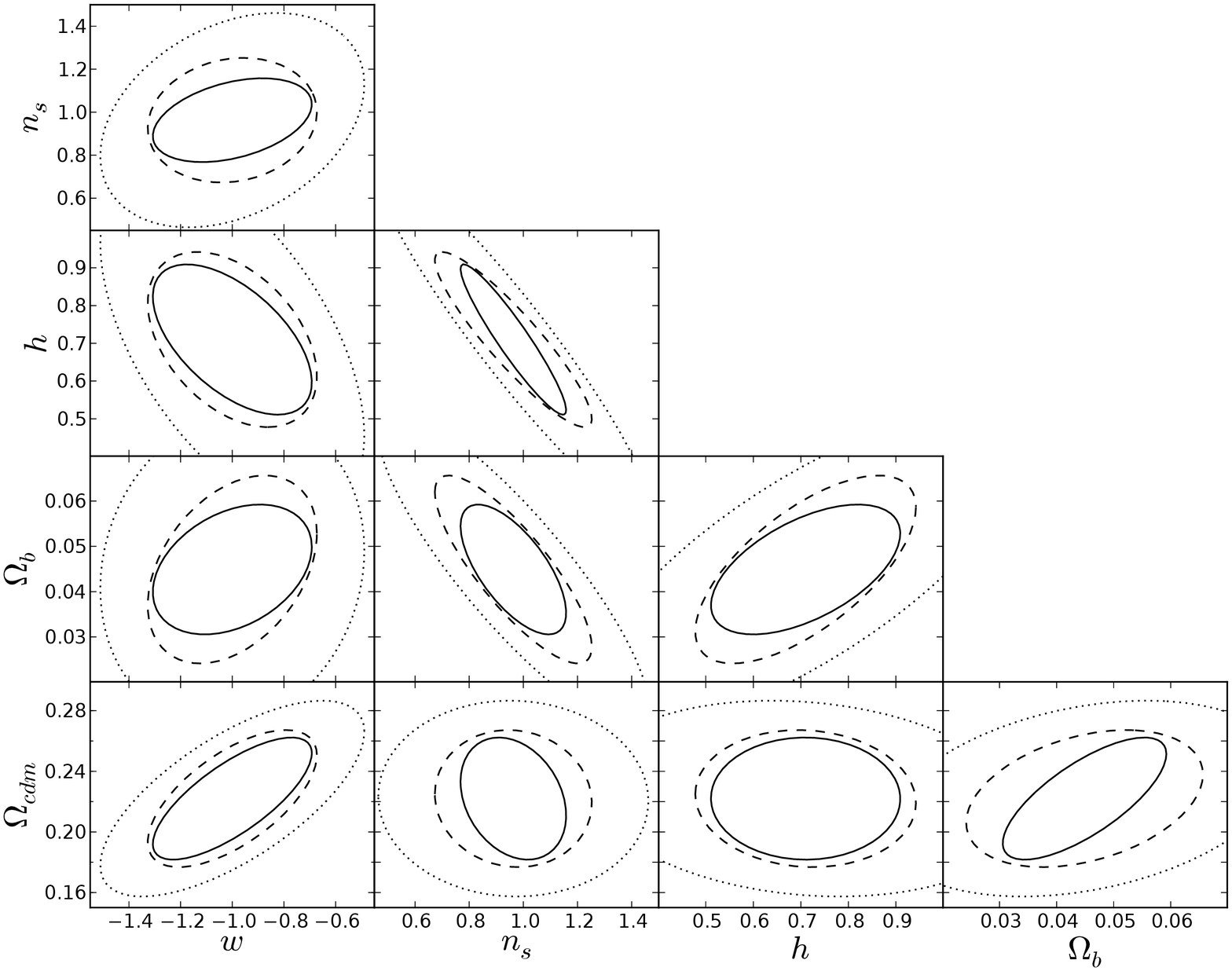}
  \end{center}
  \caption{\small 
$1 \sigma$ forecasts for $ n_s, h, \Omega_b, \Omega_{cdm}$ and $w$, 
marginalizing over $\sigma_8$ and bias,
for three scenarios: optimistic (solid line) ($20<r<200$ Mpc and $\delta_\sigma = 0.03$),
fiducial (dashed line) ($60<r<160$ Mpc and $\delta_\sigma = 0.03$) and pessimistic (dooted line) ($60<r<160$ Mpc and $\delta_\sigma = 0.05$).  
}
\label{escada}
\end{figure}

In Fig.(\ref{escada}) we plot the $1\sigma$ ellipses for the parameters 
$ n_s, h, \Omega_b, \Omega_{cdm}$ and $w$, marginalizing over $\sigma_8$ and bias.
When two of these parameters are plotted all the others are marginalized with no priors.
We study three scenarios: optimistic ($20<r<200$ $h^{-1}$Mpc and $\delta_\sigma = 0.03$),
fiducial ($60<r<160$ $h^{-1}$Mpc and $\delta_\sigma = 0.03$) and pessimistic 
($60<r<160$ $h^{-1}$Mpc and $\delta_\sigma = 0.05$).
One can see that a larger photo-z error degrades substantially the possible forecasts.
This is expected since for larger photo-z errors the signal coming from the 
spatial correlation function becomes weaker. 
The forecast error for $w$ goes from $\sim 0.3$ with $\delta_{\sigma} = 0.03$ to $\sim 0.5$ 
for $\delta_{\sigma} = 0.05$.

One can also notice from Fig.(\ref{escada}) that the constraint in $w$ is not very sensitive
to the scales probed. Nevertheless, the other parameters $\Omega_{cdm}$, $\Omega_{b}$, $n_{s}$ and $h$ are significantly dependent on the scales probed, 
especially $\Omega_{b}$ and $n_{s}$. 
This shows that those cosmological parameters change the ACF at all scales and not 
only around the BAO feature like $w$. 

%\begin{figure*}
%\epsscale{1.0} 
%\plotone{Panel_cosmo_ACF.eps} \figcaption[Panel_cosmo_ACF.eps]{ 
%$1 \sigma$ forecasts for $ n_s, h, \Omega_b, \Omega_{cdm}$ and $w$, 
%marginalizing over $\sigma_8$ and bias,
%for three scenarios: optimistic (solid line) ($20<r<200$ Mpc and $\delta_\sigma = 0.03$),
%fiducial (dashed line) ($60<r<160$ Mpc and $\delta_\sigma = 0.03$) and pessimistic (dotted line) %($60<r<160$ Mpc and $\delta_\sigma = 0.05$).  
%\label{escada}}
%\end{figure*}

\begin{figure}[h!]
  \begin{center}
    \includegraphics[width=3.in]{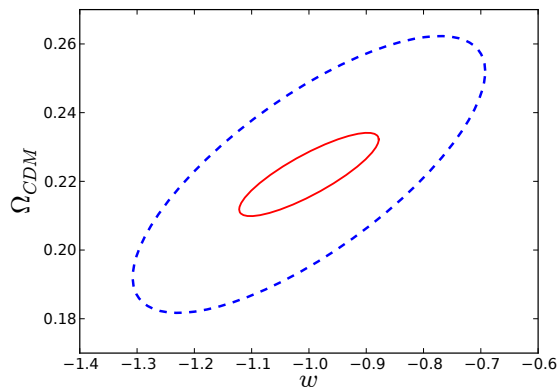}
  \end{center}
  \caption{\small 
$\Omega_{cdm} \times w$ $1 \sigma$ forecasts in the optimistic scenario in the 
case of no priors (dashed line), and adding priors from WMAP7 and HST as explained in the text
(solid line).
}
  \label{priors}
\end{figure}

Finally we investigate the role of priors in the forecasts in Fig.(\ref{priors}).  
In particular, we use the HST prior on $h$  \cite{riess09}
and WMAP7 priors from the acoustic scale $l_{A}$ , the redshift of decoupling $z_{\ast}$ 
and the shift parameter $R$ (table 10 of Komatsu et al \cite{wmap7}) and 
also imposed a prior in $n_{s}$ from WMAP7. The WMAP7 covariance matrix in 
$l_A$, $z_\ast$ and $R$ was then converted to our set of 
cosmological parameters using a method described in Mukherjee et al \cite{mukherjee2008} and 
Wang et al \cite{wang2010}.

We find the 1-$\sigma$ marginalized error on a given cosmological parameter $p$ via
\begin{equation}
\sigma_p = \sqrt{\left[F^{-1} \right]_{pp}}.
\end{equation}

In our optimistic best case scenario one would be able
to obtain $\sigma_w = 0.2$
and $\sigma_{\Omega_{cdm}} = 0.03$ without priors
and $\sigma_w = 0.08$
and $\sigma_{\Omega_{cdm}} = 0.009$ with priors.

\subsection{Impact of fiducial bias}

The analysis presented until now was performed assuming a fiducial galaxy bias $b=2$ 
for each redshift shell. This is the usual value for catalogues of luminous red galaxies (LRG).
However, it is more likely that the galaxy bias will have lower values for the complete 
galaxy set. In this subsection we show how our results change if the galaxies in the
catalogue are assumed to be unbiased on average with $b=1$, as done in \cite{ross11}.

As shown in \cite{Tegmark97}, the statistical reach of a survey can be assessed from its effective volume, which is proportional to $b^{2}$. Hence, it is expected that the constrain on the cosmological parameters will degrade if the fiducial bias decreases. 
In order to evaluate the impact of the bias on the cosmological parameters, we have 
performed the Fisher matrix analysis with a fiducial bias $b=1$ and compared to the results 
arising from $b=2$. 

\begin{figure}[h!]
  \begin{center}
    \includegraphics[width=3in]{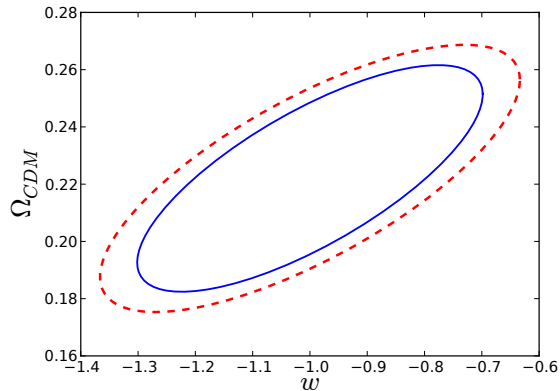}
  \end{center}
  \caption{\small $1\sigma$ forecats for $w$ and $\Omega_{cdm}$ for the optimistic scenario with two fiducial bias: $b=2$ (blue solid line) and $b=1$ (red dashed line).}
  \label{bias}
\end{figure}

In Fig.(\ref{bias}), we show the forecast for our optimistic scenario.
In this case the constrain on $w$ degrades by $17\%$, with respect to $b=2$, and the 
expected constrain on $w$ is $\sigma_{w} = 0.24$. This analysis shows that the fiducial 
galaxy bias does have a minor impact in the dark energy constraints.      

\section{Summary and Conclusions}

In this paper we have investigated the forecasts on cosmological parameters
that can be obtained from a study of the full shape of the 
2-point angular correlation function
that will be measured by future large-scale photometric surveys, 
such as the Dark Energy Survey project.
The angular correlation function was modelled taking into account
redshift space distortion, nonlinear corrections to the power spectrum, bias and
gaussian photo-z errors. After a proper binning in angle and redshift, 
the Fisher information matrix was constructed from
a covariance matrix that includes correlation of a given redshift bin with 
the 4 nearest bins. We considered 26 parameters: 
$ n_s, \sigma_8, h, \Omega_b, \Omega_{cdm}, w, b_1, b_2, \cdots, b_{20}$
with different bias for each shell.

We used the $26 \times 26$ Fisher matrix to obtain 1$\sigma$ ellipses for the 
cosmological parameters around the adopted fiducial values obtained from the central values
of WMAP7 measurements. In all our results we marginalize over the bias parameters
and $\sigma_8$. After showing results for several combinations of parameters in
Fig.(\ref{escada}), we concentrate on forecasts on the $\Omega_{cdm} \times w$ plane.
We have shown the improvements of using priors from other experiments such as WMAP7 and
HST in comparison to a simple marginalization of parameters.
Finally, we showed that the effect of neglecting the second term in the Fisher matrix
eq.(\ref{fisher}) is in fact unimportant, as usually assumed in the literature 
\cite{Tegmark97}.

In the appendix we make a comparison between the Fisher matrix approximated methodology
and a full likelihood search in parameter space which validates the Fisher matrix
approach.

We find that under these assumptions the Dark Energy Survey should be able to 
constrain the dark energy equation of state parameter $w$ and
the cold dark matter density $\Omega_{cdm}$  with a precison of 
$20 \%$ and $13 \%$ respectively  
from the full shape of the angular correlation function alone. 
When combined with priors from other observations the precision in the
determination of these parameters increase to $8 \%$ and $4 \%$ respectively.

For comparison, the WiggleZ Dark Energy Survey has recently obtained 
$w = -1.6 ^{+ 0.6}_{-0.7}$  from BAO data alone, which is improved to
$w= -0.982^{+ 0.154}_{-0.184}$ when WMPA7 distance priors are included \cite{wigglez},
which amounts to a precision of approximately $40 \%$ ($15 \%$ with priors).
Therefore, measurements of the angular correlation function from a DES-like survey 
will improve the constrain on $w$ by a factor of ${\cal O}(2)$, in the case where 
only galaxy clustering measurements are used. 
It is also expected that the constraints we have found will improve as the 
redshift range, area and number of galaxies improve with future photometric surveys.

In conclusion, our results indicate that an analysis of the full shape of the 2-point
angular correlation function can in fact bring an important contribution
to the determination of cosmological parameters in future photometric redshift surveys. 
This contribution is of course
complementary to the other observables that will be used in a DES-like survey, including 
cluster number counts,
shear from weak lensing and distances from SNIa.
One challange that remains is to combine our results with these other probes in
a consistent manner, including the proper correlations.

%\newpage
%\appendix*

\section*{Appendix}

In this Appendix we check the reliability of the Fisher matrix forecast by computing the 
likelihood function on a grid of values for the parameters.
Here we concentrate on a single redshift shell, $0.70\leq z \leq 0.75$, and examine 4 free parameters, namely $w$, $\Omega_{CDM}$, $b$ and $\sigma_{8}$, with a grid constructed around their fiducial values, see Table (\ref{grid}).
Once the grid is created it is fairly easy to search the likelihood function and compare it 
with the Fisher matrix approach. 

\begin{table}[h]
\begin{center}
\begin{tabular}{|c|c|c|c|}\hline 
 Parameter & lower bound & upper bound & $N_{grid}$  \\ \hline
 $w$ & -2.0 & 0.0 & 80  \\ \hline
 $\Omega_{CDM}$ & 0.05 & 0.55 & 100  \\  \hline
 $\sigma_{8}$ & 0.1 & 2.1 & 100  \\ \hline
 $b$ & 0.05 & 5.55 & 100 \\ \hline
\end{tabular}
\caption{\label{grid} Grid parameters.}
\end{center}
\end{table}

\begin{figure}[h!]
  \begin{center}
    \includegraphics[width=3in]{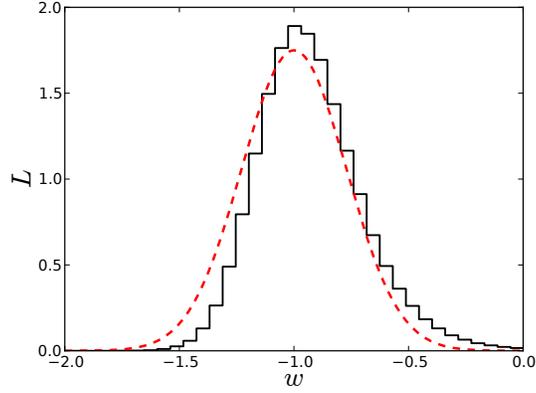}
  \end{center}
  \caption{\small The actual distribution for $w$ estimated with the grid when the other 3 free parameters are marginalized (solid line) compared with the Gaussian distribution predicted by the Fisher matrix (dashed line).}
  \label{fisherxgrid_w}
\end{figure}

In Fig.(\ref{fisherxgrid_w}) we compare the likelihood for $w$ when marginalizing 
over the other 3 parameters. 
In this case, the Fisher matrix prediction is in good agreement with the results from the grid. 
The Fisher matrix forecasts an error of  $\sigma_{w}(fisher) = 0.23$, exactly the same value obtained from the grid. 
It is interesting to note that the true distribution has a small skewness towards higher values, showing that it is not a perfect gaussian distribution, 
but in overall both distributions are in fair agreement.

\begin{figure}[h!]
  \begin{center}
    \includegraphics[width=3in]{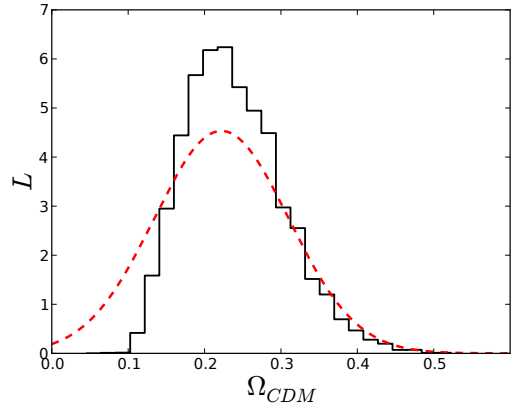}
  \end{center}
  \caption{\small The actual distribution for $\Omega_{CDM}$ estimated with the grid when the other 3 free parameters are marginalized (solid line) compared with the Gaussian distribution predicted by the Fisher matrix (dashed line). }
  \label{fisherxgrid_cdm}
\end{figure}

For $\Omega_{CDM}$ we find the same trend as for $w$, but here the 
difference between Fisher matrix and the grid likelihood is slightly larger,
as shown in Fig.(\ref{fisherxgrid_cdm}). 
In this case the $\Omega_{CDM}$ has a more pronounced non-gaussianity, with a 
skewness towards higher values. However, the prediction for the standard deviation 
from Fisher matrix is again 
in fairly good agreement with the grid, with 
$\sigma_{CDM}(fisher) = 0.088$ and $\sigma_{CDM}(grid) = 0.064$. In this case the Fisher matrix method overestimates 
the error on $\Omega_{CDM}$ due to the non-gaussianity of the likelihood. 

This simple comparison shows that the Fisher matrix gives reliable predictions for the forecasts 
of the cosmological parameters $w$ and $\Omega_{CDM}$ that we obtained in this work. 

\acknowledgments
We wish to thank Filipe Abdalla, Martin Crocce, Enrique Gazta\~{n}aga, Mariana Penna-Lima,
Ashley Ross, Sandro Vitenti, the DES-Brazil Collaboration and the
DES LSS-WG for useful comments. We especially thank Marcos Lima for a careful reading
of this manuscript and valuable suggestions.

This research was carried out with the support of the Laborat\'orio Interinstitucional de e-
Astronomia (LIneA) operated jointly by the 
Centro Brasileiro de Pesquisas F\'{\i}sicas (CBPF), the
Laborat\'orio Nacional de Computa\c{c}\~ao Cient\'{\i}fica (LNCC) and the Observat\'orio 
Nacional (ON) and
funded by the Minist\'erio de Ci\^encia e Tecnologia (MCT).

F.S. is supported by a PhD grant from CAPES.
F.dS. acknowledges financial support from the CNPq (DTI grant 381.392/09-0 associated with the PCI/MCT/ON Program).
The work of R.R. is supported by Fapesp and CNPq. 
L.N.dC. acknowledges CNPq grants
476277 / 2006 and 304.202/2008-8, FAPERJ grants E-26/102.358/2009 and E-26/110.564/2010,
and FINEP grants 01.06.0383.00 and 01.09.0298.00.
M.M. is partially supported by CNPq 
(grants 312876/2009-2, 486138/2007-0, and 312425/2006-6) 
and FAPERJ (grant E-26/171.206/2006).

%\newpage

%\newpage

%\bibliographystyle{unsrt}
\section*{References}

\end{document}